\documentclass[reprint,amsmath,amssymb,aps,floatfix]{revtex4-2}
\usepackage{graphicx}
\graphicspath{ {./Imagesv3/} }
\usepackage{dcolumn}
\usepackage{bm}
\usepackage{hyperref}
\usepackage{lipsum}
\usepackage{subcaption}  
\usepackage{caption}         
\usepackage{float}
\usepackage{tikz}
\usepackage{ragged2e}
\usepackage{placeins}
\usetikzlibrary{positioning, arrows.meta}  
\usepackage{relsize}  
\usepackage{overpic}  
\usepackage{placeins}
\usepackage{amssymb}
\usepackage{amsmath}
\usepackage{makecell}
\usepackage{bm}
\usepackage{placeins}
\usepackage[export]{adjustbox}
\usepackage{amsthm}

\theoremstyle{remark}

\usepackage{mathtools}
\usepackage{float}
\usepackage{tikz}
\usepackage{ragged2e}
\usetikzlibrary{positioning, arrows.meta}  
\usepackage{relsize}  
\usepackage{overpic}

\usepackage{tabularx}
\usepackage{booktabs}
\usepackage{array}
\begin{document}

\preprint{APS/123-QED}

\title{How Physical Dynamics Shape the Properties of Ising Machines:\\ Evaluating Oscillators vs. Bistable Latches as Ising Spins}

\author{Abir Hasan, Nikhat Khan, Nikhil Shukla}
\affiliation{University of Virginia, Charlottesville, VA, 22904 USA}

\begin{abstract}
Ising machines exploit the natural dynamics of physical systems to minimize the Ising Hamiltonian and thereby address computationally hard combinatorial optimization problems. This paradigm has motivated a range of physical implementations. In the electronic domain, coupled networks of oscillators and bistable latches have emerged as two prominent realizations of Ising machines and are the focus of the present work. Despite this common abstraction, we demonstrate that differences in the underlying physical dynamics of oscillators and latches lead to fundamentally different stability properties of the resulting dynamical systems. Specifically, we show analytically that in Bistable Latch Ising Machines (BLIMs) all discrete Ising configurations possess identical linear stability, whereas in Oscillator Ising Machines (OIMs) the Jacobian spectrum depends explicitly on the spin configuration, enabling selective destabilization of higher-energy states. We further corroborate this analysis using finite-noise perturbation experiments initialized near prescribed Ising configurations.
These results highlight how the characteristics of the device nonlinearity directly shape the local dynamical properties of Ising machine implementations.

\end{abstract}
                             
\maketitle

\section{Introduction}
The idea of exploiting the intrinsic tendency of physical systems to minimize their energy for solving computational problems finds a home in Ising machines. Such physical systems embody dynamics that aim to minimize the Ising Hamiltonian---a well-known computationally intractable problem. The Ising Hamiltonian is defined as, $H = -\sum_{i,j} J_{ij} s_i s_j $ where each spin $s \in \{+1, -1\}$, and $J_{ij}$ denotes the coupling strength between spins $i$ and $j$. Considering that many practically-relevant problems in combinatorial optimization can be expressed in terms of minimization of $H$~\cite{21_Karp_problems}, such Ising machines have the potential to enable accelerators for the broad class of combinatorial optimization problems (COPs).

A plethora of physical substrates spanning the quantum, optical, acoustic, mechanical and electronic domains have been evaluated for realizing Ising machines~\cite{mohseni2022ising,todri2024computing,Johan_Acoustic}. 
Nevertheless, not all spins are created equal, and the properties of the individual spins have a strong impact on the functional properties of the Ising machine. 
Here, we focus on the electronic implementations, specifically, using oscillators~\cite{Antik_IEDM,Khairul_JxCDC} and bistable latches (i.e., cross-coupled inverters)~\cite{BLIM_Jayjeet,Mallick2023,Hung:EECS-2024-95_BLIM} which have been commonly used to implement the underlying Ising spins. 

We investigate how the choice of the spin impacts dynamical properties such as local stability of the corresponding Ising machines, namely, Oscillator Ising Machine (OIM) and Bistable Latch Ising Machine (BLIM). We note that manifestations of BLIM include BRIM (Bistable, Resistively Coupled Ising Machines)~\cite{zhang2022cmos_BRIM,10285565_QuBRIM}. We assume resistive coupling in all our analysis.\\\\

\section{Dynamics of BLIM}
\noindent
We first focus on the dynamical properties of the BLIM and subsequently compare them to those of the Kuramoto oscillator--based Ising machine, which has been extensively studied in prior work~\cite{Khairul_Stability,Yi_Control,wang2019oim_oscillator_ising}. Starting with a single bistable latch, the normalized internal state $\nu_i(t)$ evolves according to~\cite{BLIM_Jayjeet},\\
\begin{equation}
\dot{\nu}_i
=
\frac{\tanh\!\big(k\,\tanh(k\nu_i)\big)-\nu_i}{\tau},
\label{eq:single_latch}
\end{equation}
where $\nu_i$ is the dimensionless latch state variable (with saturated values $\nu_i\approx \pm 1$ corresponding to the two logical states), $k>0$ is a dimensionless gain parameter controlling the sharpness of the nonlinearity, and $\tau$ is the intrinsic latch time constant. In an $RC$ realization, $\tau$ is proportional to an $RC$ ($R:$ Resistance $C:$ Capacitance) product (e.g., $\tau = RC$), setting the relaxation rate of an isolated latch.

\noindent
When coupled with other latches, the resulting dynamics can be expressed as,

\begin{widetext}
\begin{equation}
\begin{aligned}
\dot{\nu}_i
&=
\frac{\tanh\!\big(k\,\tanh(k\nu_i)\big)-\nu_i}{\tau}
-\left[\sum_{\substack{j=1\\ j\neq i}}^{N}
\frac{A_{ij}^{+}(\nu_i-\nu_j)}{\tau_c}
+\sum_{\substack{j=1\\ j\neq i}}^{N}
\frac{A_{ij}^{-}\!\left(\nu_i-(-\tanh(k\nu_j)\right))}{\tau_c}\right],
\label{eq:coupled_latch_assy}\\
&=
\frac{\tanh\!\big(k\,\tanh(k\nu_i)\big)-\nu_i}{\tau}
-\frac{1}{\tau_c}\left[\sum_{\substack{j=1\\ j\neq i}}^{N}
\left(A_{ij}^{+}+A_{ij}^{-}\right)\nu_i
-\sum_{\substack{j=1\\ j\neq i}}^{N}
A_{ij}^{+}\nu_j
+\sum_{\substack{j=1\\ j\neq i}}^{N}
A_{ij}^{-}\tanh(k\nu_j)
\right],
\end{aligned}
\end{equation}
\end{widetext}

\noindent
where $\tau_c$ is the coupling time constant associated with the interconnection circuitry. The positive and negative entries of the symmetric adjacency (coupling) matrix $A\in\mathbb{R}^{N\times N}$ are separated as,
\begin{equation}
A_{ij}^{+}=\max(A_{ij},0), 
\quad 
A_{ij}^{-}=\max(-A_{ij},0),
\label{eq:A_split}
\end{equation}
so that $A_{ij}^{+},A_{ij}^{-}\ge 0$, and
\begin{equation}
A = A^{+} - A^{-}.
\end{equation}
Under this decomposition, $A^{+}$ encodes ferromagnetic (attractive) interactions and $A^{-}$ encodes the \emph{magnitude} of the anti-ferromagnetic (repulsive) interactions.

Eq~\eqref{eq:coupled_latch_assy}  reveals an asymmetry in the dynamics for ferromagnetic and anti-ferromagnetic coupling in the BLIM with an additional nonlinear transformation on $\nu_j$, $\tanh(k\nu_j)$, appearing in the interaction. This feature was also noted in~\cite{Hung:EECS-2024-95_BLIM}.

\noindent Examining from an energetics standpoint, the dynamics in Eq.~\eqref{eq:coupled_latch_assy} can be considered to perform gradient descent on an energy function that can be expressed as,

\begin{widetext}
\begin{equation}
\begin{split}
E(\nu)
= \sum_{i=1}^{N} U(\nu_i)
&-\frac{1}{2\tau_c}\left[\sum_{ij}(A_{ij}^+)\nu_i\nu_j -\sum_{ij}(A_{ij}^-)\nu_i\tanh(k\nu_j)
-
\sum_i (d_i^+ + d_i^-)\nu_i^2\right].
\label{eq:modified_energy_assy}
\end{split}
\end{equation}
\end{widetext}
\noindent where $U'(\nu)=-g(\nu)$, i.e.,
\(
U(\nu)=-\int^{\nu} g(\xi)\,d\xi.
\)
Furthermore,
\(
d_i^{\pm} = \sum_{j} A_{ij}^{\pm},
\)
where $d_i^{\pm}$ denote the absolute weighted degrees associated with the positive and negative edges, respectively.

At the fixed points $\nu \in \{-1,+1\}$, the energy function in Eq.~\eqref{eq:modified_energy_assy} can be approximated to represent Ising spins, since $\tanh(k\nu_j) \approx \nu_j$ as the dynamics
settle to the saturated states $\nu \in \{-1,+1\}$ (see Appendix~\ref{app:single_node_reduction}). A sufficiently large gain $k$ ensures that the state of the nodes approaches $\nu \rightarrow \pm 1$.
In this regime, the energy function reduces to
\begin{equation}
E(\nu)\equiv H
=
C-\frac{K}{2}\left[\sum_{ij} A_{ij} s_i s_j\right],
\label{eq:modified_energy}
\end{equation}
where, 
\[
C \approxeq \sum_{i=1}^{N} U(\nu_i) 
+ \frac{1}{2\tau_c}\sum_{i}(d_i^{+}+d_i^{-})
\]
is a constant offset and $K = \tfrac{1}{\tau_c}$.

\begin{figure}
    \centering
    \includegraphics[width=0.95\linewidth]{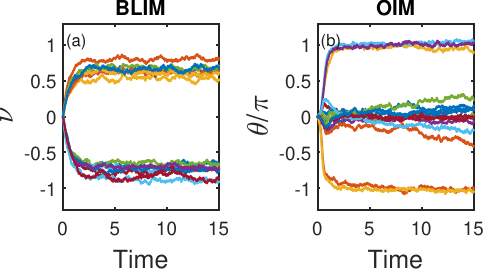}
    \caption {\justifying Temporal evolution of the state of the node in (a) BLIM, and (b) OIM, simulated on a randomly generated graph with 15 nodes and 50 edges. Simulation parameters: BLIM: $\tau=1$, $\tau_c=12$ $k=10$; $K_{\mathrm{n}}=0.05$; OIM: $K_{\mathrm{osc}}=1$; $K_s=1.5$; $K_\mathrm{n}=0.05$}
    \label{fig:BLIM_OIM_Dynamics}
\end{figure}

Figure~\ref{fig:BLIM_OIM_Dynamics}(a) shows the dynamics of a BLIM simulated for a randomly generated graph with 15 nodes and 50 negatively coupled edges. The nodes are initialized to $\nu=0$. As expected from the discussion above, $\nu$ diverges towards $\nu=-1$ or $\nu=+1$. For comparison, the dynamics of OIM (Eq.~\eqref{eq:oim_dyn}) which uses oscillators under second harmonic injection have been included as Fig.~\ref{fig:BLIM_OIM_Dynamics}(b).

\section{Linear Stability of BLIM}
We now analyze the linear stability of the BLIM dynamics and compare it with that of the OIM. In the saturated regime $\nu \rightarrow \pm 1$, corresponding to the Ising fixed points, the nonlinearity can be approximated as $\tanh(k\nu_j) \approx \nu_j$, and Eq.~\eqref{eq:coupled_latch_assy} reduces to,

\begin{widetext}
\begin{equation}
\begin{aligned}
\dot{\nu}_i
&=
\frac{\tanh\!\big(k\,\tanh(k\nu_i)\big)-\nu_i}{\tau}
-\left[\sum_{\substack{j=1\\ j\neq i}}^{N}
\frac{A_{ij}^{+}(\nu_i-\nu_j)}{\tau_c}
+\sum_{\substack{j=1\\ j\neq i}}^{N}
\frac{A_{ij}^{-}\!\left(\nu_i-(-\nu_j)\right)}{\tau_c}\right],\\\\
&=
\frac{\tanh\!\big(k\,\tanh(k\nu_i)\big)-\nu_i}{\tau}
-\frac{1}{\tau_c}\left[\sum_{\substack{j=1\\ j\neq i}}^{N}
\left(A_{ij}^{+}+A_{ij}^{-}\right)\nu_i
-\sum_{\substack{j=1\\ j\neq i}}^{N}
\left(A_{ij}^{+}-A_{ij}^{-}\right)\nu_j\right].
\end{aligned}
\label{eq:coupled_latch1}
\end{equation}
\end{widetext}

Defining the elementwise nonlinearity
$g(\nu_i)\equiv \big[\tanh\!\big(k\,\tanh(k\nu_i)\big)-\nu_i\big]/\tau$ and the
absolute weighted degree
\begin{equation}
d_i \equiv \sum_{\substack{j=1\\ j\neq i}}^{N}\left(A_{ij}^{+}+A_{ij}^{-}\right),
\qquad
D \equiv \mathrm{diag}(d_1,\ldots,d_N),
\end{equation}
Eq.~\eqref{eq:coupled_latch1} can be written compactly as,
\begin{equation}
\dot{\boldsymbol{\nu}}
=
\boldsymbol{G}(\boldsymbol{\nu})
-\frac{1}{\tau_c}\Gamma\,\boldsymbol{\nu},
\qquad
\Gamma \equiv D-(A^{+}-A^{-}),
\label{eq:coupled_latch2}
\end{equation}
where $\boldsymbol{G}(\boldsymbol{\nu})$ applies $g(\cdot)$ elementwise.

For the purely ferromagnetic case, $A_{ij}^{-}=0 \Rightarrow \Gamma\,=D-\,(A_{ij}^{+}-A_{ij}^{-})=D-A$. Consequently, $\Gamma$ then represents the \emph{Laplacian} $\mathbf{L}$ of the graph. In contrast, for the purely anti-ferromagnetic case, $A_{ij}^{+}=0 \Rightarrow \Gamma\,=D+A_{ij}^{-}=D+|A|$, which effectively represents the \emph{signless Laplacian}, $\mathbf{Q}$, of the graph.

Equation~\eqref{eq:coupled_latch1} reveals an important structural feature of the latch dynamics: each node contains a degree-dependent self-term that effectively introduces a local bias in the dynamics. Although, as seen from the energy function in Eq.~\eqref{eq:modified_energy_assy}, this term does not alter the mapping to the Ising Hamiltonian at the binary states, it modifies the local field experienced by each node during the transient evolution. 


Equation~\eqref{eq:coupled_latch1} can also be expressed as,

\begin{equation}
\dot{\nu}_i
=
g(\nu_i)
-
\frac{1}{\tau_c}
\big[
\underbrace{(d_i^+ + d_i^-)}_{\substack{\text{absolute weighted}\\ \text{degree-dependent}\\ \text{offset}}}\nu_i
-
\sum_{j}(A_{ij}^{+} - A_{ij}^-)\nu_j
\big],
\label{eq:expanded_dynamics}
\end{equation}
\noindent which shows that the dynamics of each node $i$ is influenced by a self-bias term whose influence is asymmetrically controlled by $(d_i^+ + d_i^-)$ in the general case. Comparing the above dynamics to the OIM, the degree-dependent self-bias term can be viewed as being structurally analogous to the onsite SHI term $K_s\sin(2\theta_i)$ in the OIM dynamics (Eq.~\eqref{eq:oim_dyn}), if the $K_s$ strength is made proportional to the absolute weighted degree of the individual nodes. Nevertheless, the precise functional forms are different.

\noindent Linearizing the dynamics in Eq.~\eqref{eq:coupled_latch2} about an equilibrium point $\boldsymbol{\nu}^\star$ yields the Jacobian
\begin{equation}
\begin{split}
J(\boldsymbol{\nu}^\star)
&=
\mathrm{diag}\!\big(g'(\nu_i^\star)\big)
-
\frac{1}{\tau_c}
\Gamma\\\\
&=
\mathrm{diag}\!\big(g'(\nu_i^\star)\big)
-
\frac{1}{\tau_c}
(D+|A|)
\label{eq:blim_jacobian}
\end{split}
\end{equation}

\noindent since we focus on the anti-ferromagnetic case. Here, $D$ refers to the absolute weighted degree.

\noindent The off-diagonal entries of the Jacobian can be expressed as,
\begin{equation}
J_{ij}
=
-\frac{1}{\tau_c}|A_{ij}|,
\qquad i\neq j,
\label{eq:blim_offdiag}
\end{equation}
which are \emph{independent} of the operating point. All state dependence enters exclusively through the diagonal terms,
\begin{equation}
J_{ii}
=
g'(\nu_i^\star)
-
\frac{d_i}{\tau_c}.
\end{equation}
where, $d_i$ refers to the absolute unweighted degree of the $i^{th}$ node.

\noindent We now restrict attention to Ising configurations,
$\nu_i^\star = s_i \in \{-1,+1\}$. Here, we assume that $k$ is large enough to approximate $s$ as fixed points. Owing to the symmetry of the latch nonlinearity, $g'(+1)=g'(-1)$, implying
\begin{equation}
\mathrm{diag}\!\big(g'(s_i)\big)
=
g'(1)\,I,
\end{equation}
where $I$ refers to an identity matrix. Consequently, at every Ising configuration,
\begin{equation}
J
=
g'(1)I
-
\frac{1}{\tau_c}
\left(
D + |A|
\right),
\label{eq:blim_jacobian_ising}
\end{equation}
which is independent of the specific spin assignment $\boldsymbol{s}$.

\noindent The maximal real eigenvalue therefore satisfies
\begin{equation}
\lambda_{\max}
=
g'(1)
-
\frac{1}{\tau_c}
\lambda_{\min}\!\left(D + |A|\right),
\label{eq:blim_lambda}
\end{equation}
and is identical for all Ising configurations of a given graph. Linear stability of all discrete states, $\nu \in \{-1,+1\}$, is thus uniform.

We note that the configuration-independent stability result is not specific to the particular latch nonlinearity used here. Rather, it relies on the symmetry of the two saturated latch states and on the coupling contribution being state-independent in the reduced model. Furthermore, the result could also be extended to other state-independent coupling implementations, including effective capacitive coupling schemes, provided that their reduced small-signal behavior can be expressed through a configuration-independent effective coupling time constant, \(\tau_c\).

\subsection*{Comparison with OIMs} 

In contrast to BLIM, OIMs which are governed by phase dynamics of the form:
\begin{equation}
\dot{\theta}_i
=
-K_{\mathrm{osc}}\sum_j A_{ij}\sin(\theta_i-\theta_j)
-
K_s\sin(2\theta_i),
\label{eq:oim_dyn}
\end{equation}
exhibit a fundamentally different linearization structure. Linearizing Eq.~\eqref{eq:oim_dyn} about an Ising configuration $\theta_i^\star \in \{0,\pi\}$ yields off-diagonal Jacobian entries
\begin{equation}
J_{ij}
=
K_{\mathrm{osc}} A_{ij}\cos(\theta_i^\star-\theta_j^\star),
\qquad i\neq j.
\end{equation}

\noindent Since
\(
\cos(\theta_i^\star-\theta_j^\star)
=
s_i s_j,
\)
the Jacobian explicitly depends on the spin configuration. Consequently, the spectrum of the Jacobian varies across Ising states and Ising energy. A detailed analysis of OIM linear stability has been reported in prior work~\cite{Yi_Control}. Furthermore, Allibhoy~\textit{et al.}~\cite{allibhoy2025global} have shown that the conditional expectation of the eigenvalues of the Jacobian decreases with reducing Ising energy making them more likely to be stabilized.
\begin{figure}[!h]
\centering
\includegraphics[width=0.9\columnwidth]{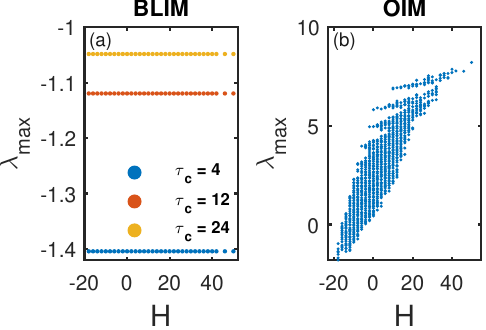}
\caption{\justifying Comparison of maximal Jacobian eigenvalues ($\lambda_{\mathrm{max}}$) across all Ising configurations for (a) the BLIM and (b) the OIM. The same graph as in Fig. 1 is considered. In the BLIM, $\lambda_{\max}$ is identical for all configurations at fixed $\tau_c$, indicating configuration-independent linear stability. In the OIM, $\lambda_{\max}$ depends explicitly on the spin configuration, reflecting configuration-dependent stability. $k=2$ (BLIM) and $K_s=1.5$ (OIM).}
\label{fig:BLIM_vs_OIM}
\end{figure}

We also expect the OIM stability results to extend qualitatively to parametrically driven oscillator platforms such as coherent Ising machines (CIMs) operating in near-threshold weak-coupling regimes and the amplitude saturates rapidly to an approximately homogeneous value. In this case, the dynamics can be reduced to a Kuramoto-style phase description~\cite{khan2026analyzing}. When amplitude dynamics are strong and there is amplitude heterogeneity, the effective couplings can be rescaled as \(\widetilde{W}_{ij}=r_i r_j W_{ij}\), and a full amplitude--phase stability analysis would be required.

This distinction between the linear stability properties of the BLIM and the OIM is illustrated in Fig.~\ref{fig:BLIM_vs_OIM} using the same graph considered in Fig.~\ref{fig:BLIM_OIM_Dynamics}. For the BLIM (Fig.~\ref{fig:BLIM_vs_OIM}(a)), the maximal eigenvalue $\lambda_{\max}$ remains constant across all the Ising configurations, and therefore is independent of the Ising energy, for a fixed $\tau_c$. Consequently, this produces \emph{horizontal bands} when plotted against Ising energy $H$. Here, $\lambda_{\max}$ denotes the largest eigenvalue of the Jacobian; $\lambda_{\max}<0$ indicates asymptotic stability of the corresponding configuration, whereas $\lambda_{\max}>0$ indicates instability. In contrast, the OIM (Fig.~\ref{fig:BLIM_vs_OIM}(b)) exhibits a broad distribution of $\lambda_{\max}$ values that depend on the Ising energy, reflecting configuration-dependent linear stability.\\
\begin{figure*}
    \centering
\includegraphics[width=0.9\linewidth]{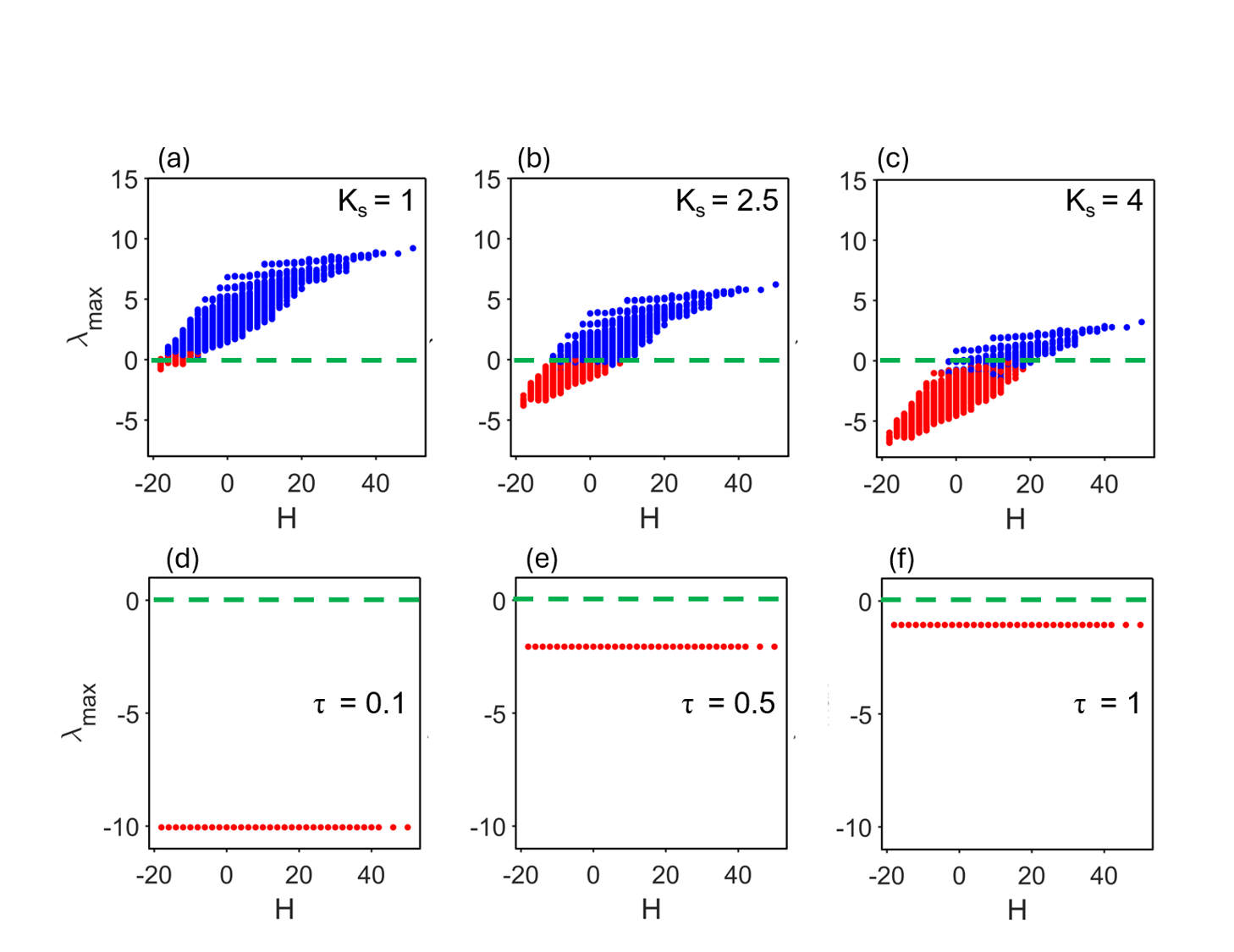}
\caption{\justifying Finite-noise probe of local stability near Ising configurations for a representative 15-node graph. For each binary configuration, the system is initialized near the corresponding Ising state and evolved under weak noise. 
Red points denote configurations that remain in the same binary state, while blue points denote configurations that escape to a different binary state. 
(a)--(c) OIM results for \(K_s=1,\,2.5,\,4\), respectively. 
(d)--(f) BLIM results for \(\tau=0.1,\,0.5,\,1\), respectively. 
The horizontal axis is the Ising energy \(H\), the vertical axis is the largest Jacobian eigenvalue \(\lambda_{\max}\). 
Simulation parameters: noise amplitude \(A_n=0.1\), OIM coupling \(K=1\), BLIM gain \(k=10\), and BLIM coupling time constant \(\tau_c=30\).}
\label{fig:Benchmark}
\end{figure*}

To directly test the local-stability implication of the Jacobian analysis, we perturb the dynamics in the vicinity of the binary Ising configurations. The results have been presented in Fig.~\ref{fig:Benchmark}. For the same representative 15-node graph considered in Fig.~\ref{fig:BLIM_OIM_Dynamics}, we enumerate all \(2^{N-1}=2^{14}\) distinct Ising configurations \(\mathbf{s}\in\{\pm 1\}^{N}\), where configurations related by a global spin reversal, \(\mathbf{s}\) and \(-\mathbf{s}\), are treated as equivalent. For each configuration, the BLIM is initialized near \(\nu_i=s_i\), while the OIM is initialized near the corresponding phase state \(\theta_i=0\) for \(s_i=+1\) and \(\theta_i=\pi\) for \(s_i=-1\). We then add weak noise and integrate the dynamics for a fixed observation window. We interpret the response to the perturbation as a finite-noise probe of local stability: configurations that remain in the same binary state after the noisy evolution are taken to remain within the local basin of attraction (red points), whereas configurations that switch to a different binary state are interpreted as having escaped that basin under perturbation (blue points).

 For the OIM, the largest Jacobian eigenvalue varies strongly across Ising configurations and is generally correlated with the Ising energy. The blue points are concentrated among configurations with \(\lambda_{\max}>0\), consistent with the prediction that these configurations are unstable under the linearized dynamics. Increasing \(K_s\) shifts the spectrum toward more negative values, thereby stabilizing a larger fraction of configurations.

In contrast, for the BLIM, \(\lambda_{\max}\) is independent of the Ising configuration and remains negative. Thus, all binary Ising configurations are locally asymptotically stable, irrespective of their Ising energy. This is
reflected in the perturbation experiment by the red points across all configurations: the initialized binary states remain stable under finite-noise perturbations. The noise-perturbation test therefore supports the central local-stability distinction between the two dynamical systems. A corresponding analysis for the \(N=800\) \(G1\)--\(G5\) instances from the G-set benchmark \cite{Gset_HelmbergRendl_Stanford} is provided in Appendix~\ref{ap:Gset}, where the same trend is observed.

It is important to note that the above analysis characterizes the local dynamics in the vicinity of states that can be mapped to Ising configurations. It therefore does not describe the full phase space of the dynamics, including fixed points at intermediate latch amplitudes that do not correspond directly to saturated binary configurations. Consequently, these results should not be interpreted as a complete description of the overall dynamical behavior or optimization performance of the two systems. Both latch-based~\cite{sharma2022increasing} and oscillator-based~\cite{moy20221,bashar2020experimental} Ising-machine demonstrations have reported promising performance in hardware implementations, underscoring that local stability near saturated binary states is only one factor governing practical solver performance.

\section{Conclusion}
In this work, we elucidate fundamental differences in the dynamical properties of BLIMs and OIMs arising from differences in the underlying nonlinearity of the spin. Oscillator-based implementations exhibit configuration-dependent linear stability at the discrete Ising configurations, whereas bistable latch-based Ising machines exhibit configuration-independent linear stability across these states. This distinction implies that, in OIMs, the stability of a binary state depends on the corresponding Ising configuration and energy, while in BLIMs all binary configurations share the same local stability structure.  We further corroborate this Jacobian-based picture through finite-noise perturbation experiments initialized near prescribed Ising configurations. Our results highlight how the physical nonlinearity of the underlying spin element shapes the local dynamical properties of Ising machine implementations.

Besides the differences in dynamical properties, it is also important to note that other factors such as the ease of hardware implementation will also play an important role in the choice of the dynamical platform. BLIMs are attractive because the spin element can be realized using compact CMOS latch structures, with switching speeds set primarily by local RC time constants and without requiring explicit phase generation or injection locking. OIMs generally require additional oscillator, coupling, and synchronization circuitry, which can increase area, energy cost, and sensitivity to device mismatch or phase noise. Thus, the eventual choice between BLIMs and OIMs will depend on how circuit-level constraints such as area, energy, speed, and noise tolerance are balanced along with the distinct dynamical properties of each platform.

\section*{Acknowledgement}
\noindent This material is based upon work supported by ARO award W911NF-24-1-0228.
{
\section*{Data availability}
\noindent The data supporting the findings of this study are publicly available on GitHub at \\ \nolinkurl{https://github.com/AbirHasan092/OIM_BLIM_code}\\[0.8em]
\section*{Code Availability}
\noindent The codes associated
with this manuscript are publicly available on GitHub at \\ \nolinkurl{https://github.com/AbirHasan092/OIM_BLIM_code}\\[0.8em]

\clearpage

\appendix
\section{Reduction of the Cross-Coupled Latch Dynamics to a Single Differential State}
\label{app:single_node_reduction}

Here, we demonstrate that under mild symmetry conditions the bistable latch dynamics collapse onto an invariant manifold, permitting the approximation $\nu_i' \approx -\nu_i$ after the initial transient dynamics.

Consider a symmetric cross-coupled inverter pair described by
\begin{equation}
\dot{\nu}_i = \frac{g(\nu_i') - \nu_i}{\tau},
\qquad
\dot{\nu}_i' = \frac{g(\nu_i) - \nu_i'}{\tau},
\label{eq:latch_nodes}
\end{equation}
where $\nu_i$ and $\nu_i'$ denote the normalized node voltages on the two complementary sides of the latch, $\tau$ is the intrinsic latch time constant, and $g(\cdot)$ represents the static inverter transfer characteristic. In the present work,
\begin{equation}
g(\nu) = \tanh\!\big(k\,\tanh(k\nu)\big),
\end{equation}
which is an odd, saturating nonlinearity.

To analyze the symmetry of the dynamics, we define differential and common-mode variables
\begin{equation}
\mu_i \equiv \frac{\nu_i - \nu_i'}{2},
\qquad
\varepsilon_i \equiv \frac{\nu_i + \nu_i'}{2}.
\label{eq:mode_def}
\end{equation}

Expressing the node voltages in terms of these variables,
\begin{equation}
\nu_i = \varepsilon_i + \mu_i,
\qquad
\nu_i' = \varepsilon_i - \mu_i .
\end{equation}

Substituting these expressions into Eq.~\eqref{eq:latch_nodes} yields
\begin{equation}
\begin{aligned}
\dot{\mu}_i
&=
\frac{1}{2\tau}
\Big[
g(\varepsilon_i - \mu_i) - g(\varepsilon_i + \mu_i)
- 2\mu_i
\Big],
\\
\dot{\varepsilon}_i
&=
\frac{1}{2\tau}
\Big[
g(\varepsilon_i - \mu_i) + g(\varepsilon_i + \mu_i)
- 2\varepsilon_i
\Big].
\end{aligned}
\label{eq:s_dyn}
\end{equation}

Since $g(\cdot)$ is odd,
\begin{equation}
g(-\nu) = -g(\nu).
\end{equation}

Evaluating Eq.~\eqref{eq:s_dyn} on the subspace $\varepsilon_i = 0$ gives
\begin{align}
\dot{\mu}_i
&=
\frac{g(\mu_i) - \mu_i}{\tau},
\label{eq:s_reduced}
\\
\dot{\varepsilon}_i
&=
-\frac{\varepsilon_i}{\tau}.
\label{eq:c_linear}
\end{align}

Equation~\eqref{eq:c_linear} shows that the common-mode variable $\varepsilon_i$ decays exponentially with rate $1/\tau$. Consequently, the subspace $\varepsilon_i = 0$ is an invariant and linearly stable manifold of the dynamics. Any initial common-mode perturbation relaxes toward zero on the intrinsic latch time scale.

On this invariant manifold the dynamics reduce exactly to
\begin{equation}
\dot{\mu}_i = \frac{g(\mu_i) - \mu_i}{\tau}.
\label{eq:single_node_latch}
\end{equation}

Using $\nu_i = \varepsilon_i + \mu_i$ and $\nu_i' = \varepsilon_i - \mu_i$, and noting that $\varepsilon_i \rightarrow 0$, we obtain
\begin{equation}
\nu_i' \approx -\nu_i.
\end{equation}

Starting from the symmetric initialization $\nu_i = 0$, this approximation becomes increasingly accurate as the system evolves and the latch states approach the saturated values $\nu_i \rightarrow \pm 1$.

\section{Impact of coupling strength on BLIM dynamics}
\label{ap:BLIM_K}

In this section, we investigate the impact of coupling strength and latch gain on the dynamics of the BLIM.

\emph{Simulation framework}: Figure~\ref{fig:BLIM_characteristics} shows simulation results for a BLIM implementing a 15-node graph with 50 edges (same as the graph considered in Fig.~\ref{fig:BLIM_OIM_Dynamics}). Simulations were performed in MATLAB\textsuperscript{\textregistered} using the built-in stochastic differential equation (SDE) solver based on the Euler--Maruyama method. The simulated SDE with additive white Gaussian noise (AWGN) is
\begin{equation}
d\boldsymbol{\nu}
=
\left(\boldsymbol{G}(\boldsymbol{\nu})
-\frac{1}{\tau_c}\Gamma\,\boldsymbol{\nu}\right)dt
+K_{\mathrm{n}} I\, dW_t,
\label{eq:coupled_latch_sde}
\end{equation}
where $K_{\mathrm{n}} = 0.05$, $I$ is the identity matrix, and $dW_t$ denotes a Wiener process. The spin state is extracted as $s=\operatorname{sign}(\nu)$. All simulations shown in Fig.~\ref{fig:BLIM_characteristics} are initialized at $\nu=0$. The framework considered here is the same as that employed to generate the results in Fig.~\ref{fig:BLIM_OIM_Dynamics}(a).

Figures~\ref{fig:BLIM_characteristics}(a)--(d) show representative time evolution of the latch states for increasing $\tau_c$, which inversely controls the effective coupling strength i.e., larger $\tau_c$ corresponds to weaker coupling. The corresponding equilibrium energy distributions over 100 independent trials are shown in Fig.~\ref{fig:BLIM_characteristics}(e)--(h).

As $\tau_c$ increases, we observe that the probability of reaching the ground state decreases and higher-energy configurations become more prevalent. For the example graph considered here, the dynamics converge to the ground state ($H=-18$) in 35 out of 100 trials for $\tau_c=4$, whereas no ground-state solutions are observed for $\tau_c=24$ or $\tau_c=100$.

Figures~\ref{fig:BLIM_characteristics}(i)--(l) show representative trajectories for increasing latch gain $k$ (with $\tau_c$ fixed), and Figs.~\ref{fig:BLIM_characteristics}(m)--(p) show the corresponding equilibrium energy distributions over 100 independent trials. A similar trend is observed: increasing $k$ reduces the probability of attaining the ground state and increases the likelihood of convergence to higher-energy configurations. 

\begin{figure*}
\centering

\includegraphics[width=0.9\linewidth]{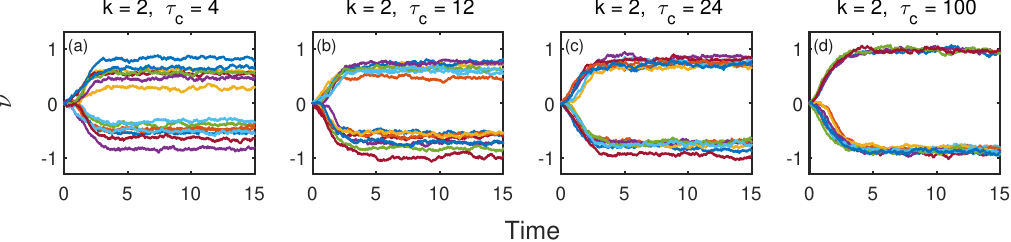}

\vspace{0.8cm}

\includegraphics[width=0.9\linewidth]{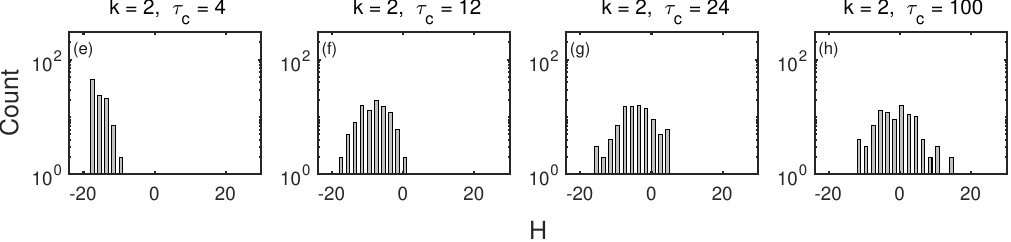}

\vspace{0.8cm}

\includegraphics[width=0.9\linewidth]{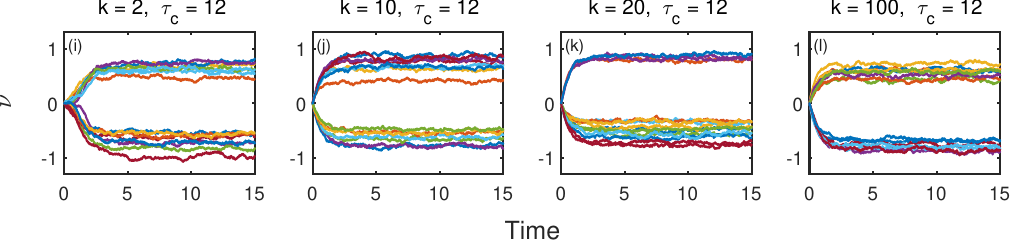}

\vspace{0.8cm}

\includegraphics[width=0.9\linewidth]{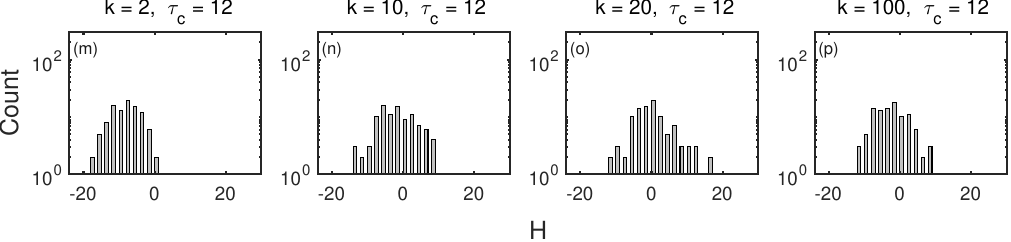}
\caption{\justifying Influence of $\tau_c$ and $k$ on BLIM dynamics. (a)–(d) Representative time evolution of node states for increasing $\tau_c$ (with fixed $k=2$). (e)–(h) Histograms of the Ising energy of configurations obtained at equilibrium over 100 independent trials for the corresponding $\tau_c$ values. (i)–(l) Time evolution of node states for increasing gain $k$ (with fixed $\tau_c=12$). (m)–(p) Energy histograms obtained over 100 independent trials for the corresponding $k$ values.}
\label{fig:BLIM_characteristics}
\end{figure*}

\begin{figure*}
    
\centering

\includegraphics[width=1.96\columnwidth]{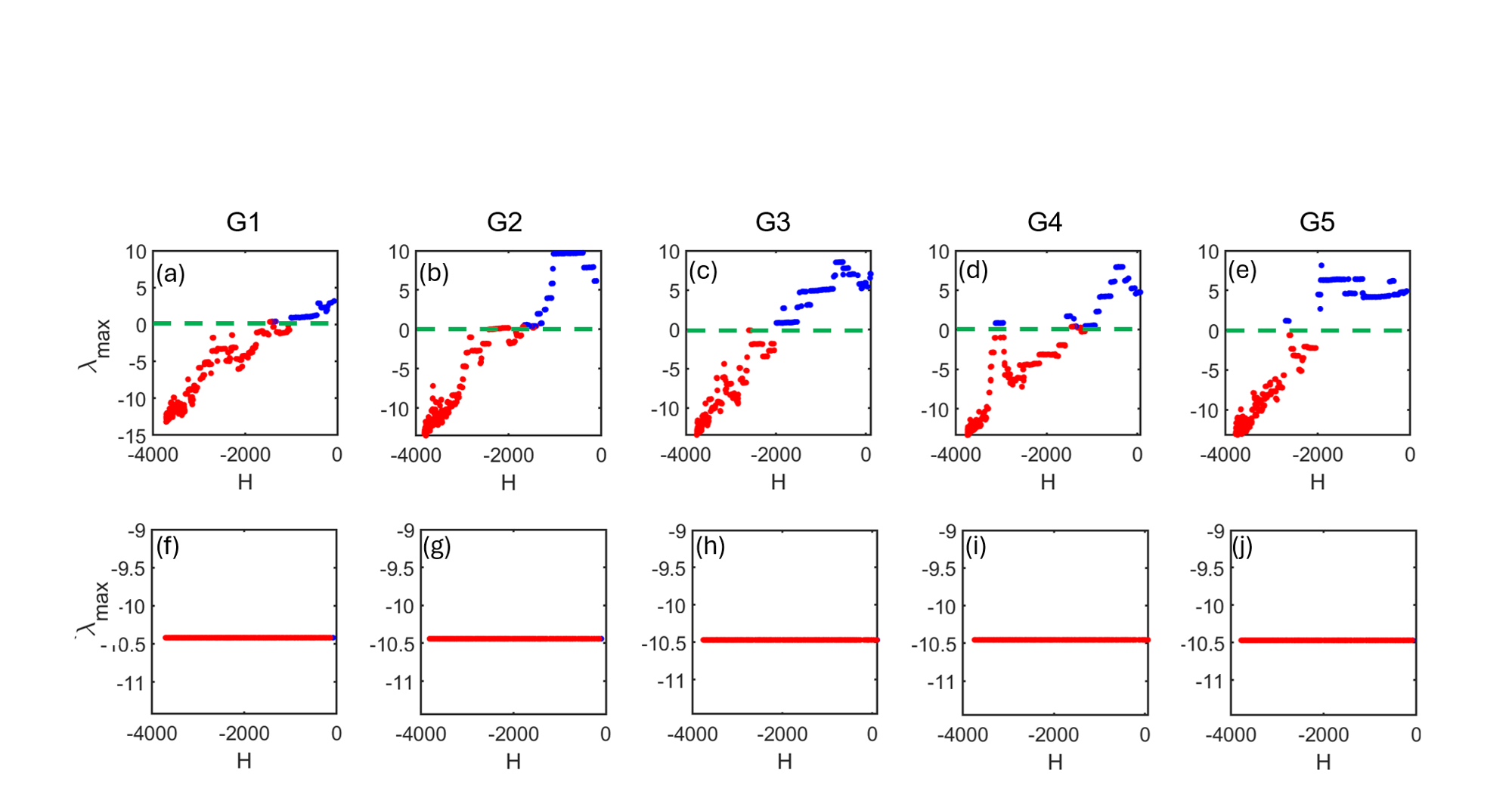}

\caption{\justifying Finite-noise probe of local stability for $2\times10^4$ sampled Ising configurations from the G1--G5 instances of the G-set benchmark. 
For each sampled binary configuration, the system is initialized near the corresponding Ising state and evolved under weak noise. Red points denote configurations that remain in the same binary state, while blue points denote configurations that escape to a different binary state. The top row shows the OIM results, for which the largest Jacobian eigenvalue \(\lambda_{\max}\) varies across sampled configurations and depends on the Ising energy \(H\). The bottom row shows the corresponding BLIM results, for which \(\lambda_{\max}\) is independent of the sampled Ising configuration. The horizontal axis is the Ising energy \(H\), the vertical axis is \(\lambda_{\max}\), and the green dashed line marks \(\lambda_{\max}=0\). }
\label{fig:BLIM_vs_OIM_GSET}
\end{figure*}

\vspace{25cm}

\section{Finite-Noise Local Stability Analysis on the G1-G5 Instances from G-set Benchmark}
\label{ap:Gset}

Here, we test the local-stability implications of the BLIM and OIM on larger \(N=800\) instances from the G-set benchmark, specifically G1--G5. Since enumerating all configurations is intractable for these graphs, we sample \(2\times 10^4\) Ising configurations using Markov Chain Monte Carlo. For each sampled configuration, we evaluate the dominant Jacobian eigenvalue and also perform a finite-noise probe of local stability by initializing the dynamics near the corresponding binary state, adding weak noise, and integrating over a fixed observation window. Configurations that remain in the same binary state are interpreted as remaining within the local basin of attraction, whereas configurations that switch are interpreted as escaping under perturbation. As shown in Fig.~\ref{fig:BLIM_vs_OIM_GSET}, the OIM exhibits energy-dependent local stability: low-energy configurations have more negative dominant eigenvalues and remain stable under weak perturbations, while higher-energy configurations approach or cross the stability boundary and are more likely to escape. In contrast, the BLIM shows an energy-independent dominant eigenvalue across the sampled configurations, indicating that the local stability of its binary states does not intrinsically distinguish low- and high-energy configurations.

\clearpage

\def\bibsection{\section*{References}}  
\bibliography{Arxiv_v2.bbl}

\end{document}